\documentclass[conference]{IEEEtran}
\IEEEoverridecommandlockouts
\usepackage[sorting=none]{biblatex}
\usepackage{amsmath,amssymb,amsfonts}
\usepackage{algorithmic}
\usepackage{graphicx}
\usepackage[english]{babel}
\usepackage{textcomp}
\usepackage{xcolor}
\definecolor{codegreen}{rgb}{0,0.6,0}
\definecolor{codegray}{rgb}{0.5,0.5,0.5}
\definecolor{codepurple}{rgb}{0.58,0,0.82}
\usepackage{hyperref}
\usepackage[all]{hypcap}
\usepackage{setspace}
\usepackage{etoolbox}
\usepackage{listings, lstautogobble}
\usepackage{url}
\usepackage{fixltx2e}
\usepackage{svg}

\lstdefinestyle{mystyle}{
	literate=*{=}{{\bfseries\textcolor{codepurple}{=}}}{1}
		{==}{{\bfseries\textcolor{codepurple}{==}}}{2}
		{*}{{\bfseries\textcolor{codepurple}{*}}}{1}
		{+}{{\bfseries\textcolor{codepurple}{+}}}{1}
		{<}{{\bfseries\textcolor{codepurple}{<}}}{1}
		{>}{{\bfseries\textcolor{codepurple}{>}}}{1}
		{>=}{{\bfseries\textcolor{codepurple}{>=}}}{2}
		{<=}{{\bfseries\textcolor{codepurple}{<=}}}{2}
		{True}{{\bfseries\textcolor{codegreen}{True}}}{4}
		{False}{{\bfseries\textcolor{codegreen}{False}}}{5}
		{1}{{\textcolor{codegreen}{1}}}{1},
	commentstyle=\color{teal},
	keywordstyle=\bfseries\color{codegreen},
	numberstyle=\tiny\color{codegray},
	stringstyle=\color{red},
	basicstyle=\ttfamily\footnotesize,
	breakatwhitespace=false,         
	breaklines=true,                 
	captionpos=b,                    
	keepspaces=true, 
	showspaces=false,                
	showstringspaces=false,
	showtabs=false,                  
	tabsize=2,
	numbers=left,
	numbersep=5pt, 
	xleftmargin=2em,
	framexleftmargin=1.5em,
	escapeinside=||
}
\let\origthelstnumber\thelstnumber
\makeatletter
\newcommand*\Suppressnumber{%
	\lst@AddToHook{OnNewLine}{%
		\let\thelstnumber\relax%
		\advance\c@lstnumber-\@ne\relax%
	}%
}
\newcommand*\Reactivatenumber[1]{%
	\setcounter{lstnumber}{\numexpr#1-1\relax}
	\lst@AddToHook{OnNewLine}{%
		\let\thelstnumber\origthelstnumber%
		\refstepcounter{lstnumber}
	}%
}
\makeatother

\AtBeginEnvironment{quote}{\singlespacing\small}
\AtEndEnvironment{quote}{\bigskip}

\def\BibTeX{{\rm B\kern-.05em{\sc i\kern-.025em b}\kern-.08em
		T\kern-.1667em\lower.7ex\hbox{E}\kern-.125emX}}
	
\providecommand{\keywords}[1]
{
	\small	
	\textbf{\textit{Keywords}:} #1
}
\AtBeginBibliography{\small}
\bibliography{main.bib}

\begin{document}
	
	\title{Analysis of Patterns in Recorded Signals of Software Systems With a Variance Based Segmentation Algorithm\\
	}

	\author{\IEEEauthorblockN{Bojan Lukić, Thorben Knust, Andreas Rausch}
	\IEEEauthorblockA{\textit{Clausthal University of Technology}\\
	\textit{Institute for Software and Systems Engineering}}\\
	}

	\maketitle
	
	\begin{abstract}
	Due to the increasing complexity and interconnectedness of different components in modern automotive software systems there is a great number of interactions between these system components and their environment.
	These interactions result in unique temporal behaviors we call underlying scenarios. The signal data from all system components, which is recorded during runtime, can be processed and analyzed by observing changes in their runtime. Different system behaviors can be characterized by dividing the whole data spectrum into appropriate segments with consistent behavior, classifying these segments, and mapping them to different scenarios. 
	These disjunctive scenarios can be analyzed for their specific behavior which may divert from the expected average system behavior.
	We state the emerging problem of data segmentation as follows: divide a multivariate data set into a suitable amount of segments with consistent internal behavior.
	The problem can be divided into 2 subproblems: "How many segments are present in the data set?", and "What are good segmentation indices for the underlying segments?".
	The complexity of the problem still needs to be assessed, however, at this point we expect it to be NP-hard, as both the number of segments and the segmentation points are unknown.
	We are in search of appropriate metrics to quantify the quality of a given segmentation of a whole data set. In this paper, we discuss the segmentation of multivariate data, but not the classification of segments into scenario classes.
	In the following, we investigate segmentation algorithms for solving the subproblem of finding suitable segmentation indices by constant amount of segments.
	The algorithms are investigated towards effectivity and efficiency by applying them to a data set taken out of a real system trace provided by our automotive partners. 
	The supplementary resources that include the data set and custom algorithms can be found in the GitHub repository\footnote{See \url{https://github.com/Bojan-Lukic/signal-analysis-with-variance-based-segmentation} for supplementary material and custom algorithms used in this paper.}.
	\end{abstract}

	\keywords{multivariate time series,	time series segmentation, signal analysis, runtime analysis of automotive signals.}
	\bigskip

	\section{Introduction}
	Nowadays, automotive software systems have different behaviors, depending on their inner and outer context, due to a high system complexity.
	Inner context describes the state based functionality that changes over time (e.g. implemented by internal variables that are periodically read and written by the same component). One example can be a component that executes a bus acknowledge every nth execution).
	Outer context describes every dependency to the environment or neighboring system. As an example one can think of a sensor signal or bus communication. 
	Behavior is defined by the reaction of the system towards this inner and outer context. 
	The reaction can be understood both as the functional reaction (i.e. controlling a system actor) and the timing behavior (i.e. how long the system takes to calculate its periodical tasks).
	
	In this work, the focus will be on the timing behavior, although the approach may as well be applicable to functional behavior. With the possibility to achieve consistent traces on software system level, the collection of all available system traces will be used as the base data for this work. 
	The aforementioned system behaviors can be characterized by dividing the whole data spectrum into appropriate segments with consistent behavior, classifying these segments and mapping them to different scenarios. Each scenario can be analyzed separately to its expected runtime behavior which in turn can be used for gaining a better system understanding, identification of optimization candidates, and assuring the system's overall real time processing capability \cite{knust}.
	\bigskip

	\section{Related Work}
	Though the work of analyzing control functions in automotive systems at a low level is just emerging, there have been efforts of signal segmentation in more common fields, such as electroencephalogram (EEG) signal analysis, in the past years.
	
	One of the papers closely related to this work comes from Henrik Peters, Falk Howar, and Andreas Rausch with their work \cite{2016Peters} on inferring environment models for control functions from recorded signal data. In their work, they infer automata models for analyzing recorded signal data from automotive systems after road tests. The conclusion is that additionally to road tests, software systems need to be analyzed by recorded data afterwards by means of segmentation of signals and classification of found segments.
	
	Aleš Procházka et al. use signal de-noising, evaluation of principal components, and segmentation based upon feature detection \cite{2010Prochazka} for analyzing EEG signals. In their paper, the authors explain the necessity of diagnostic tools for analyzing brain activity with EEG signal processing and propose the segmentation of signals with subsequent segmentation and feature extraction as one method to achieve this analysis.
	
	One more work from the domain of signal analysis via segmentation is the paper \cite{2019Perslev} from Mathias Perslev et al. which discusses the analysis of sleep data with a convolutional neural network. Their method of segmenting time series outperforms state-of-the-art deep learning models and shows high accuracy and robustness for reliably and precisely segmenting and preparing time series for further analysis. 
	\bigskip

	\section{Motivation}	
	Sought after is a method that splits a time ordered system trace containing both signal- and process-runtime information into an appropriate amount of disjunctive segments on dedicated segmentation indices. Each segment shall show a \textit{consistent behavior}.
	As there is no standardized understanding for consistent behavior and system scenarios, we define general metrics to quantify and express the goodness of a segmentation, and thus make the quality of a segmentation verifiable. By using general and not system-specific metrics, we do not include domain knowledge but derive all actual occurred scenarios out of the real data.
	
	For better understanding, we show an example with a data set consisting of only one signal:
	the problem statement can be expressed as finding a number of segments in a signal so that within each segment consistent behavior can be observed. Using a test signal, one possible segmentation can be retrieved with intuition:
	
	\begin{figure}[!htbp]
		\centerline{\includegraphics[width=\linewidth]{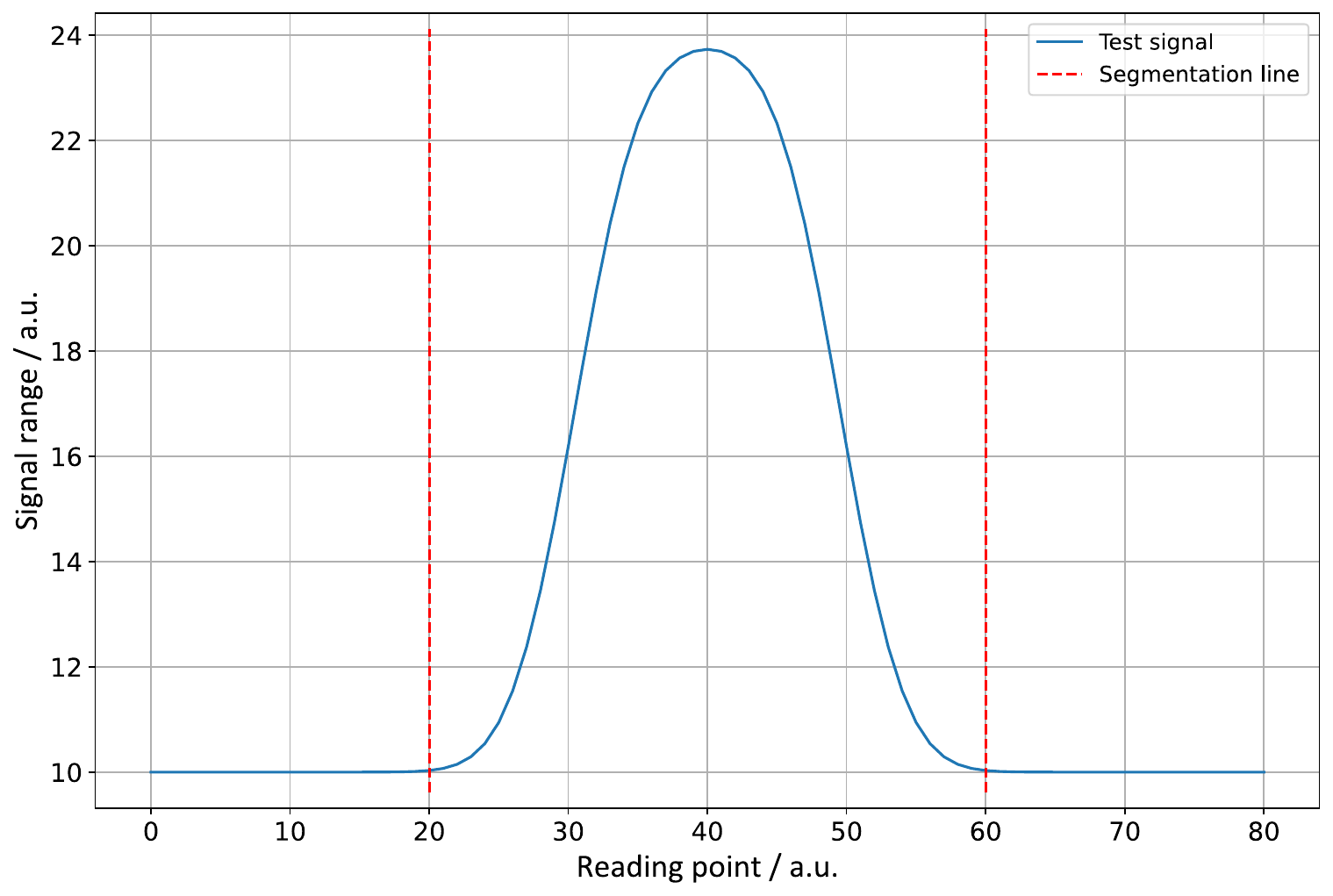}}
		\caption{Intuitive solution for a n = 3 segmentation of the test signal.}
		\label{fig:figureI}
	\end{figure}
	
	In figure \ref{fig:figureI} there are three segments defined between the start, the segmentation lines, and the end of the time series.
	The idea behind above segmentation is that the first and third segment contain parts of the signal which have a relatively low fluctuation and stay in the signal range 10. The second segment contains a part of the signal which first rises and then drops smoothly. 
	
\begin{figure}[!htbp]
		\centerline{\includegraphics[width=\linewidth]{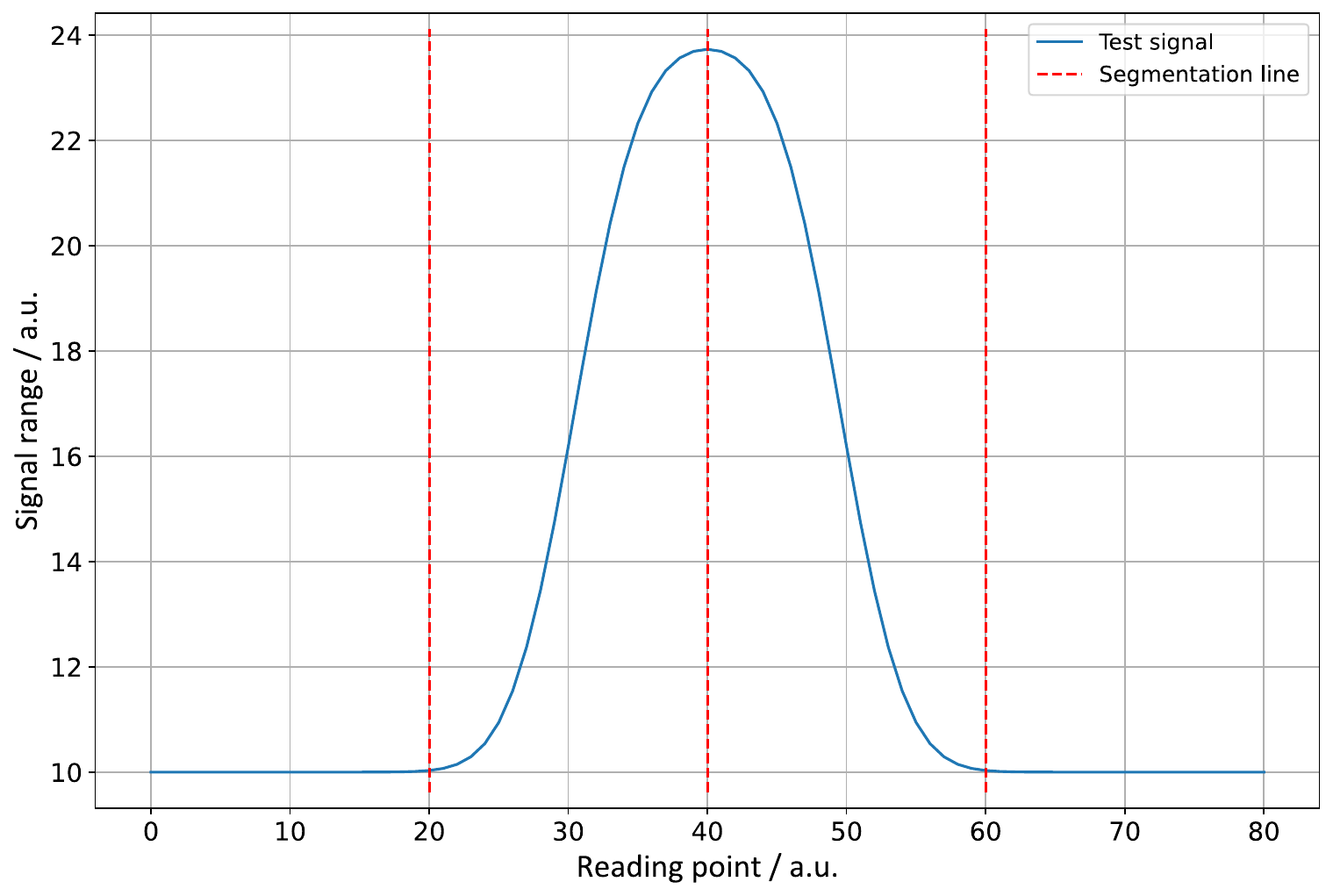}}
		\caption{Intuitive solution for a n = 4 scenario segmentation of the test signal.}
		\label{fig:figureII}
	\end{figure}

	Another possible segmentation can be seen in figure \ref{fig:figureII}.
	
	Here, there are four underlying segments. The middle segment has been divided into two smaller segments with the idea being that the rise and fall of a signal can be seen as individual and independent behaviors making for two separate segments. 
	
	Without further information it is not clear which of the presented segmentation is better. The segmentation quality cannot be measured as there is no metric at hand measuring the consistency. The developed metric in this work can be seen as an effective candidate for finding consistency in a signal, however, it is not guaranteed that this metric is the best or if there are better ones for finding consistent behavior in a signal. 
	
	Assuming the existence of a proper metric showing which segmentation is optimal, the problem can be expressed as an optimization problem consisting of two different subproblems:

	\begin{itemize}
	\item How many segments are present in the multivariate data set?
	
	One can argue that finding fewer segments with the same internal consistency towards the chosen metric should be preferred. This follows the principle of Occam's Razor\footnote{See Rasmussen, C., \& Ghahramani, Z. (2001). Occam's Razor. In T. Leen, T. Dietterich, \& V. Tresp (Reds), Advances in Neural Information Processing Systems (Vol 13).} and leads to the goal of minimizing the overall amount of segments.
	
	Applied to the example presented above, the three segment scenario would be favored over the scenario with four segments, assuming same values for the segmentation metric.
	
	\item What are the optimal segmentation indices for the underlying segments?
	
	With the given metric and a certain amount of segments, the division of a data set into individually consistent parts, the precise seperation of segments, as well as valid indices need to be chosen.
	\end{itemize}
	
	Finding an exact solution to above discussed subproblem of finding optimal and valid segmentation indices for a predefined number of segments requires trying out every combination of segmentation indices for the number of segments in the data set. This leads to a complexity of $\mathcal{O}(n^{m - 1})$ for the optimal solution in the big O notation with n being the number of samples in the data set and m being the predefined number of segments. The complexity can be reduced to $\mathcal{O}(n \cdot (m - 1))$ for an approximate solution, which will be discussed in more detail in section V.
	
	In our interpretation the two subproblems of minimizing the segments and finding segments with consistent inner behavior are working against each other: purely minimizing the number of segments in the data set would result in one segment over all data points. Segmenting with the aim of increasing consistency of segments will lead to segments consisting of single data points and eventually getting as many segments as there are data points. The optimal solution for the whole problem will therefore be a trade off between these two optimization problems. The central question is how this trade off of finding an optimal segmentation in the scope of these two optimization problems can be achieved. This question remains open, as for now no algorithm has been found which can optimally segment a data set with above criteria. This leads to the search for algorithms which deliver an approximate solution to the problem.
	Heuristics for evaluating the effectiveness of the algorithms to approximate optimal solutions will be investigated in later sections.
	\bigskip

	\section{Data model}
	In this section the used data will be presented and appropriate filtering methods for retrieving the most relevant data will be applied.
	\bigskip
	
		\subsection{Presentation of the data}
\begin{figure}[!htbp]
			\centerline{\includegraphics[width=\linewidth]{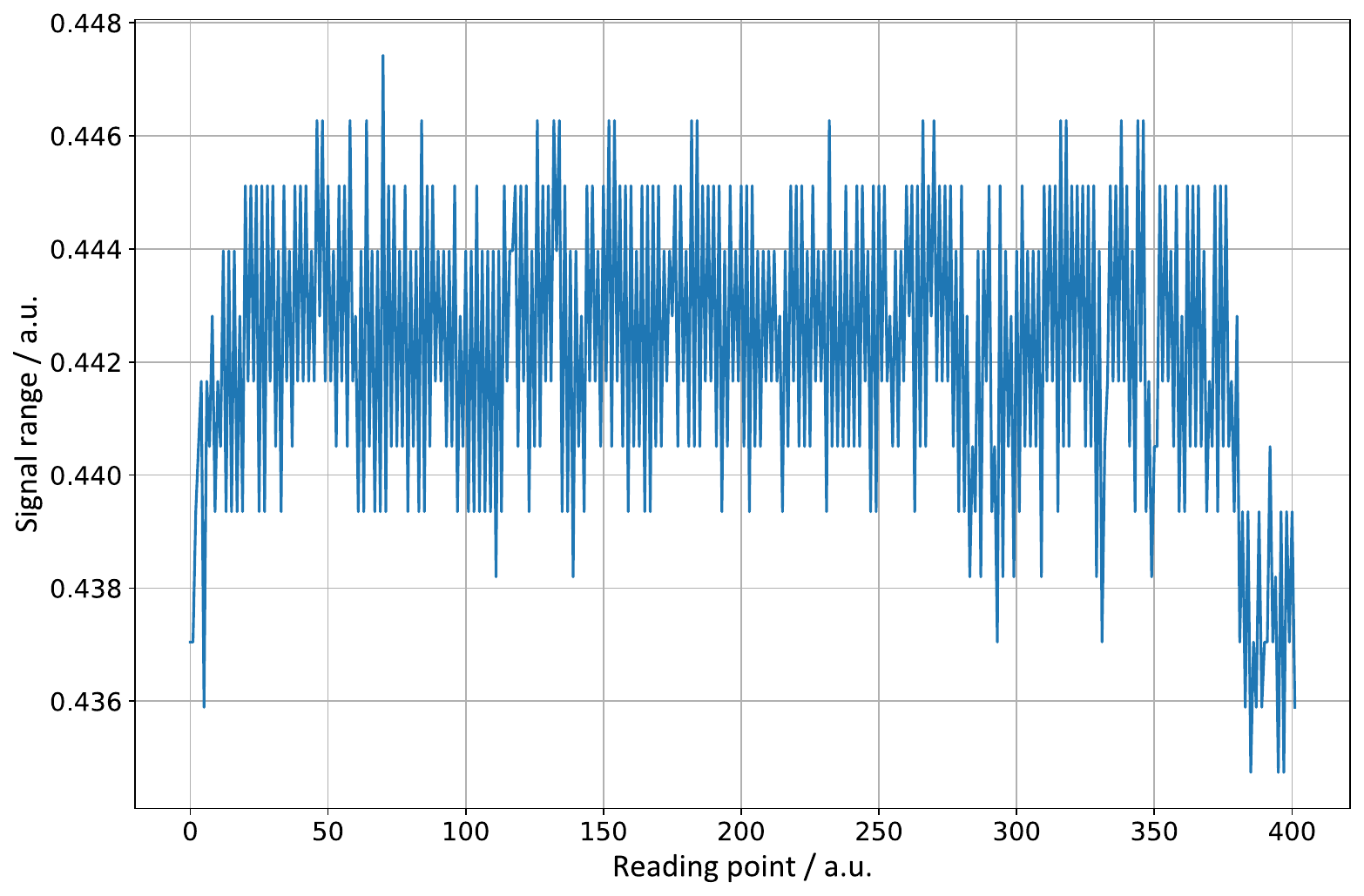}}
			\caption{Sample process contained in data set.}
			\label{fig:figureIII}
		\end{figure}

		The data used in this work consists of a matrix with 1300 rows and 402 columns. The rows represent signals or the runtime processes and the columns represent the samples or runtime values. An illustration of a signal and a process can be seen in figure \ref{fig:figureIII}, whereas figure \ref{fig:figureIV} shows the data used in this work.
	
		\bigskip
		
		\subsection{Data preparation}
\begin{figure}[!htbp]
			\centerline{\includegraphics[width=\linewidth]{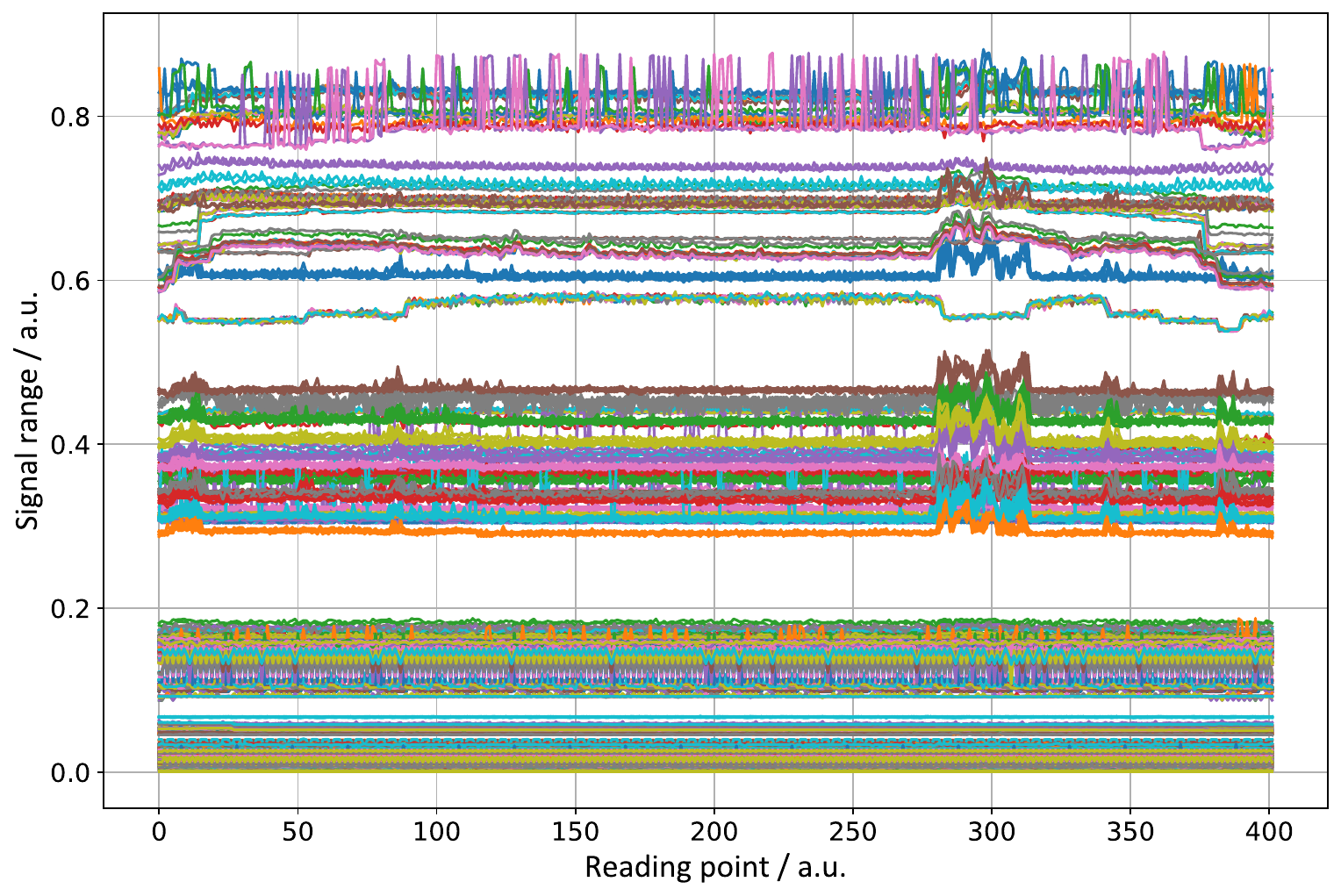}}
			\caption{Plot of all processes contained in data set.}
			\label{fig:figureIV}
		\end{figure}

		Figure \ref{fig:figureIV} shows all processes in one plot.
		
\begin{figure}[!htbp]
			\centerline{\includegraphics[width=\linewidth]{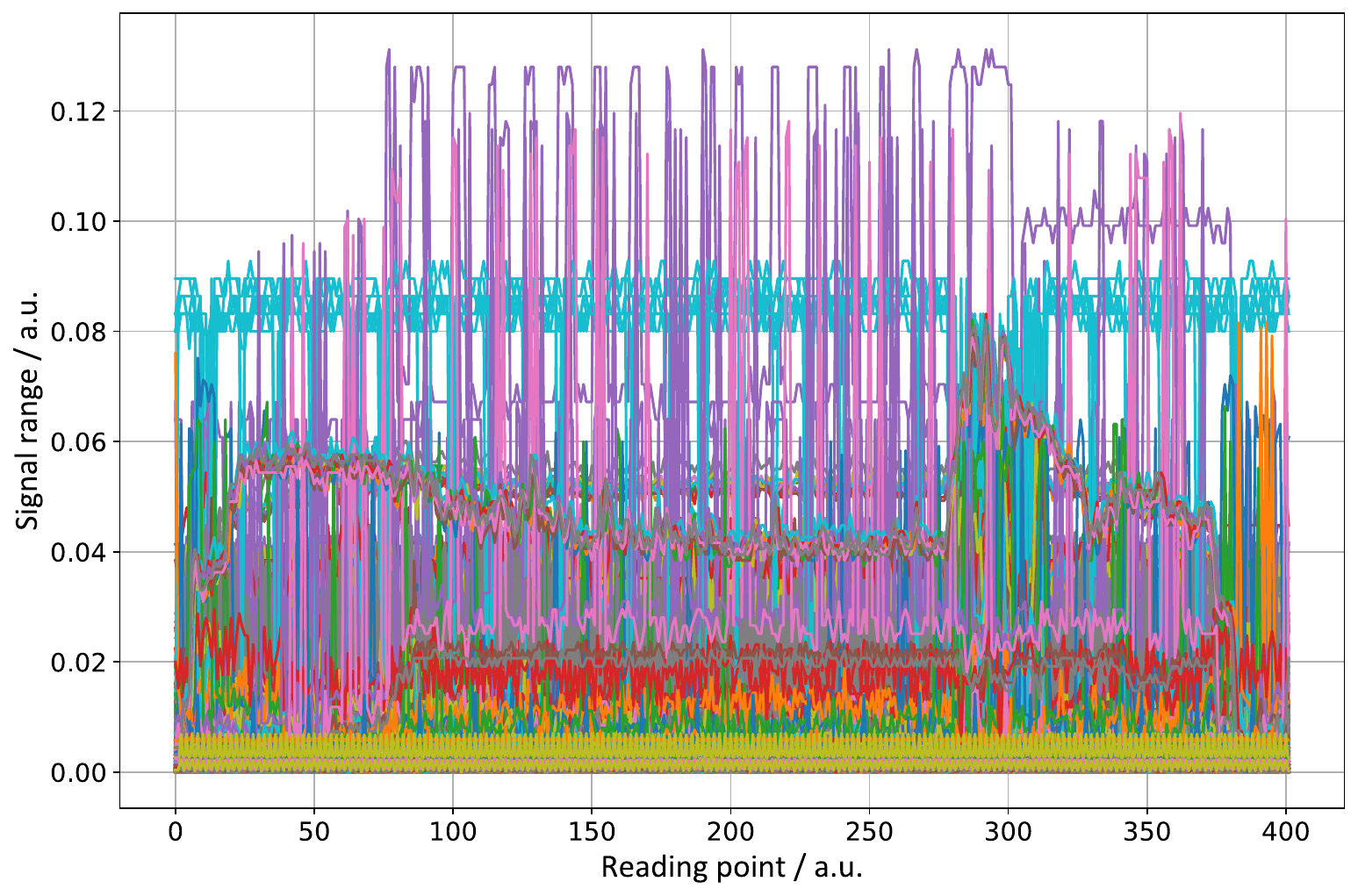}}
			\caption{Plot of all processes transformed by offset reduction.}
			\label{fig:figureV}
		\end{figure}

		For better visualization, the data set is filtered to retrieve a subset of the original data set. The aim of the filtering is to have the minimum values of all processes at a signal range of 0 and select the n processes with the highest peak-to-valley ratio. Figure \ref{fig:figureV} shows the result of the offset reduction.
		More precisely, the offset reduction is done by subtracting a constant from each data point in a process row p\textsubscript{n}. The constant c\textsubscript{p} is equal to the minimum value in p and is subtracted from each data point in p. This can be expressed as follows:
		
		\begin{equation*}
			\begin{aligned}
				&c\textsubscript{p\textsubscript{n}} = min(p\textsubscript{n})\\
				&p\textsubscript{n}\textsuperscript{new} = p\textsubscript{n} - c\textsubscript{p\textsubscript{n}}
			\end{aligned}
		\end{equation*}
		
		After applying above transformation to each process, all processes have their minimum value at signal range 0. Figure \ref{fig:figureV} shows the result of the transformation.
		
\begin{figure}[!htbp]
			\centerline{\includegraphics[width=\linewidth]{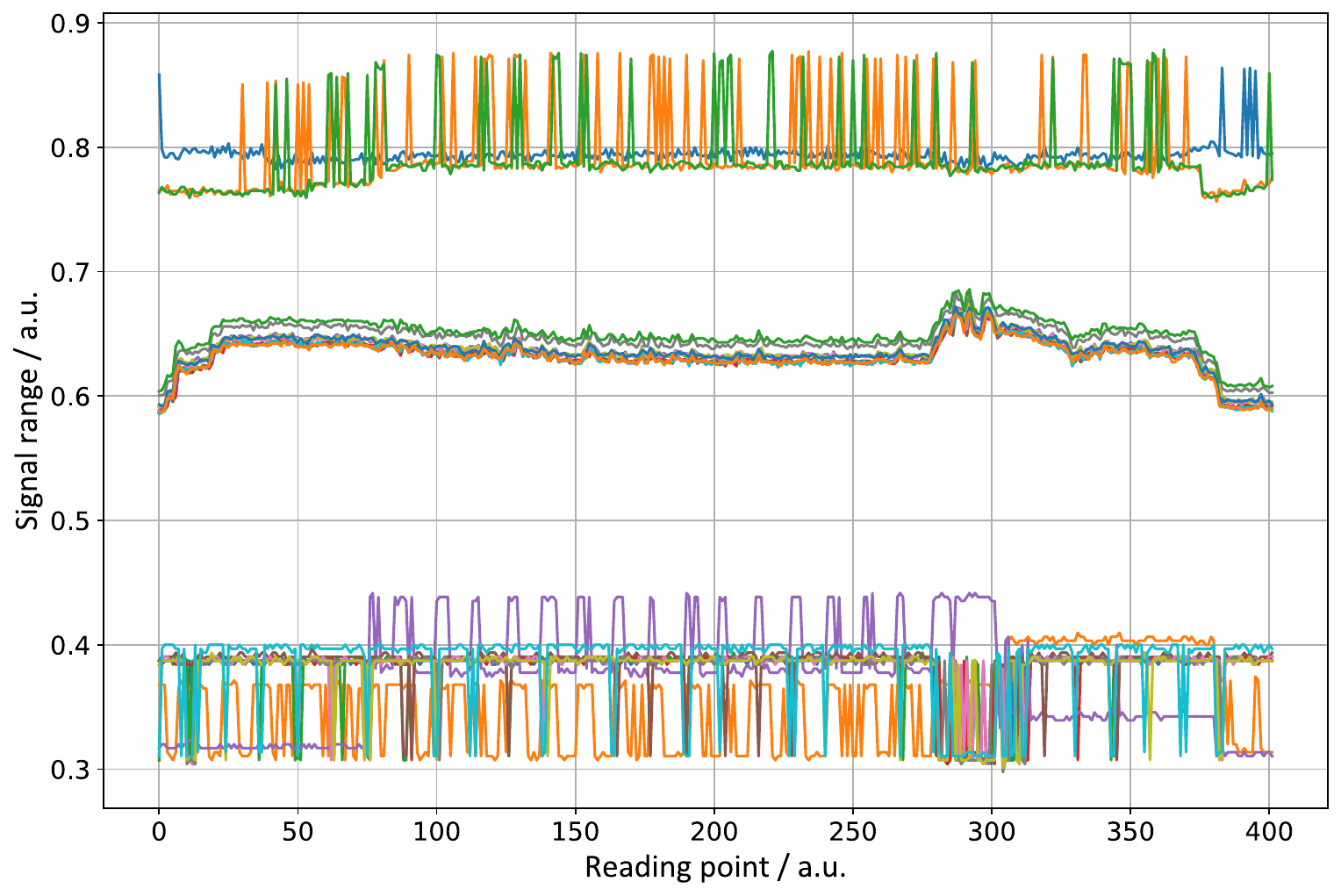}}
			\caption{Top processes by peak-to-valley ratio.}
			\label{fig:figureVI}
		\end{figure}

		Finally, the processes with the highest peak-to-valley ratio can be selected. In this case, an arbitrary number of processes with a peak-to-valley ratio higher than 60 \% of the highest peak-to-valley ratio are filtered and kept for further analysis. After applying the filtering method, out of more than 900 processes 23 processes are left and put back to their original signal range. The final result after filtering can be observed in figure \ref{fig:figureVI}.
	
		\bigskip

	\section{Segmentation with variance}
	Various methods for processing and segmenting the data set were tested. Fourier transformations, step detection algorithms, and polynomial approximations of the processes are not feasible, as the underlying data is discrete and the resulting transformations are inaccurate. 
	Slope analysis or methods such as derivation are not recommended, as segments that belong together could potentially be separated using these transformations. Another disadvantage of slope analysis is the fact that choosing the right zero points for a number of segments in many cases is not straightforward and requires either manual input or additional metrics.
	
	A more flexible and reliable method of finding segments is the minimization of the variance of segments in the data set. The calculation of the variance and the algorithm used for this work will be discussed in the next subsections.
	\bigskip
	
		\subsection{Variance of a data set}
		The	variance of a random variable is the expected value of the squared deviation from the mean of the variable:
		
		\begin{equation*}
			Var(X) = E[(X - \mu)^2] ,
		\end{equation*}
		with X being the random variable, $\mu$ the mean of random variable X, and E being the expected value.
		
		For a data set containing multiple rows as well as columns, the variance calculated for each data point per row is summed up:
		
		\begin{equation*}
			Var(M) = \sum_{i=1}^{m} \sum_{j=1}^{n} \frac{(X_{ij} - \mu_i)^2}{n} ,
		\end{equation*}
		with M being the data set containing multiple rows and columns, m the number of rows in the data set, n the number of data points in data set M, X$_{ij}$ the jth data point in ith row of data set M, and $\mu_i$ the mean of row i in data set M.
		
		For data with multiple columns and rows, as well as multiple segments, the variance is calculated for each segment per row:
		
		\begin{equation*}
			Var(M) = \sum_{h=1}^{s} \sum_{i=1}^{m} \sum_{j=1}^{n_h} \frac{(X_{ij}^h - \mu_i^h)^2}{n_h} ,
		\end{equation*}
		with s being the number of segments, n$_h$ the number of data points in segment h, X$_{ij}^h$ the jth data point in ith row of segment h, and $\mu_i^h$ the mean of row i of segment h in data set M.
		
		The goal of the segmentation algorithm is to find the segmentation indices for a specific number of segments so that the sum of variances for all segments is minimized. This problem can be expressed as an optimization problem.
		\bigskip
		
		\subsection{Optimization problem}
		As shortly discussed above, the value of the sum of variances for each segment should be minimized, making this an optimization problem. The optimization problem can be denoted as following:
		
		Given: a data set M, an amount of segments S, and a function Var: M$_{mxn}$ x s $\rightarrow$ $\mathbb{R}$ mapping data set M$_{mxn}$ with given amount of segments S to a value in $\mathbb{R}$ representing its whole variance.\\
		Sought: separators x $\in$ M such that x$_h$ $<$ x$_{h+1}$ for all S$_i$ with i $\in$ {1, ..., s - 1} and such that 
		\begin{equation*}
			\text{min } Var(M) = \sum_{h=1}^{s} \sum_{i=1}^{m} \sum_{j=1}^{n_h} \frac{(X_{ij}^h - \mu_i^h)^2}{x_h - x_{h-1}} ,
		\end{equation*}
		with the separators x being the upper segmentation lines for all segments S. The function \textit{Var(M)} should be minimized here.
		\bigskip
		
		\subsection{Brute force calculation of optimal solution}
\begin{figure}[!htbp]
			\centerline{\includegraphics[width=\linewidth]{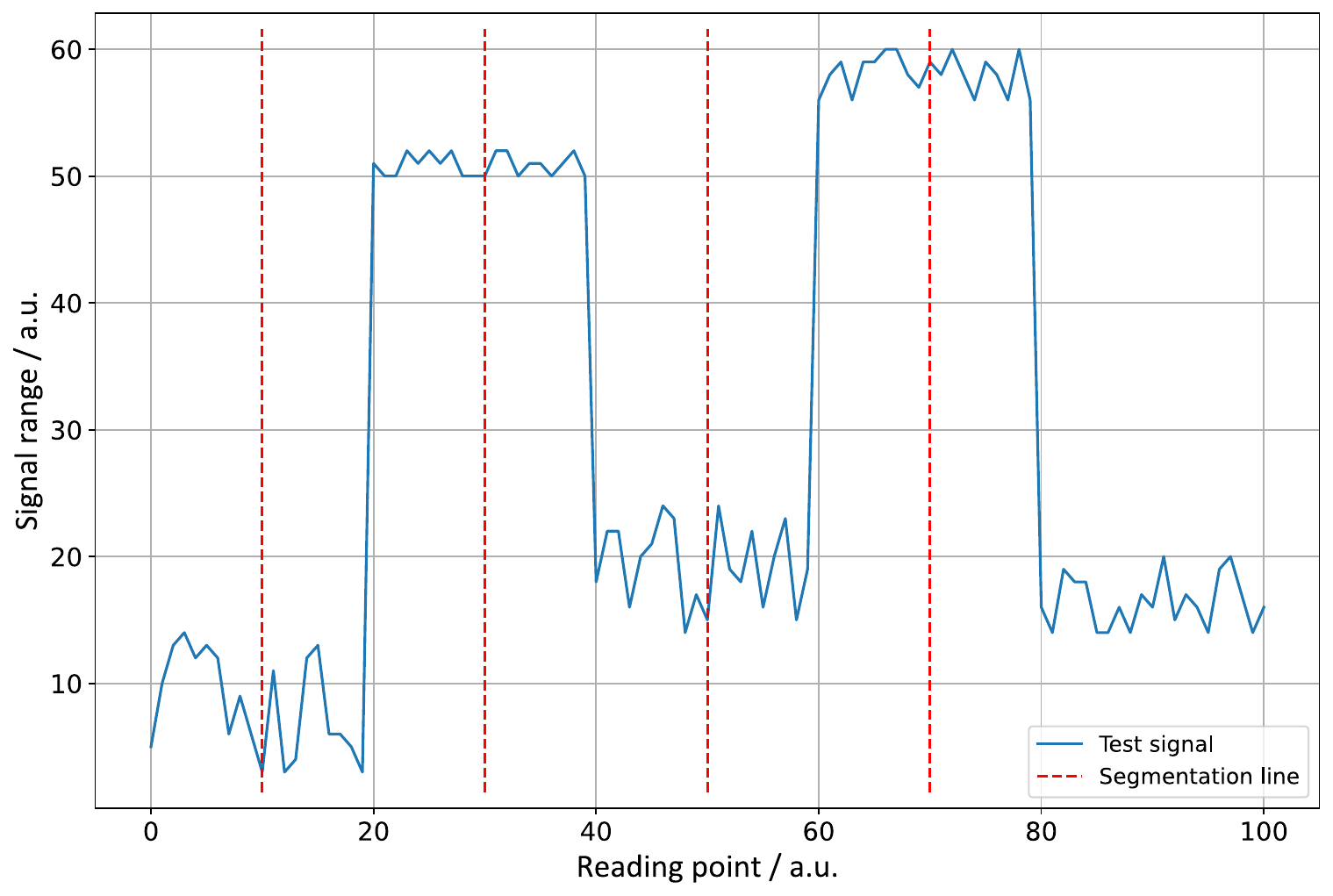}}
			\caption{Test signal with an arbitrary presegmentation.}
			\label{fig:figureVII}
		\end{figure}

		To compare further work of segmenting the processes with a custom algorithm, the optimal solution for a test signal is calculated. Here, a test signal containing 100 data points is used to calculate the optimal segmentation with 5 segments based on minimizing the value of the sum of the variances of all segments. Figure \ref{fig:figureVII} shows the test signal with an arbitrary presegmentation.
		
\begin{figure}[!htbp]
			\centerline{\includegraphics[width=\linewidth]{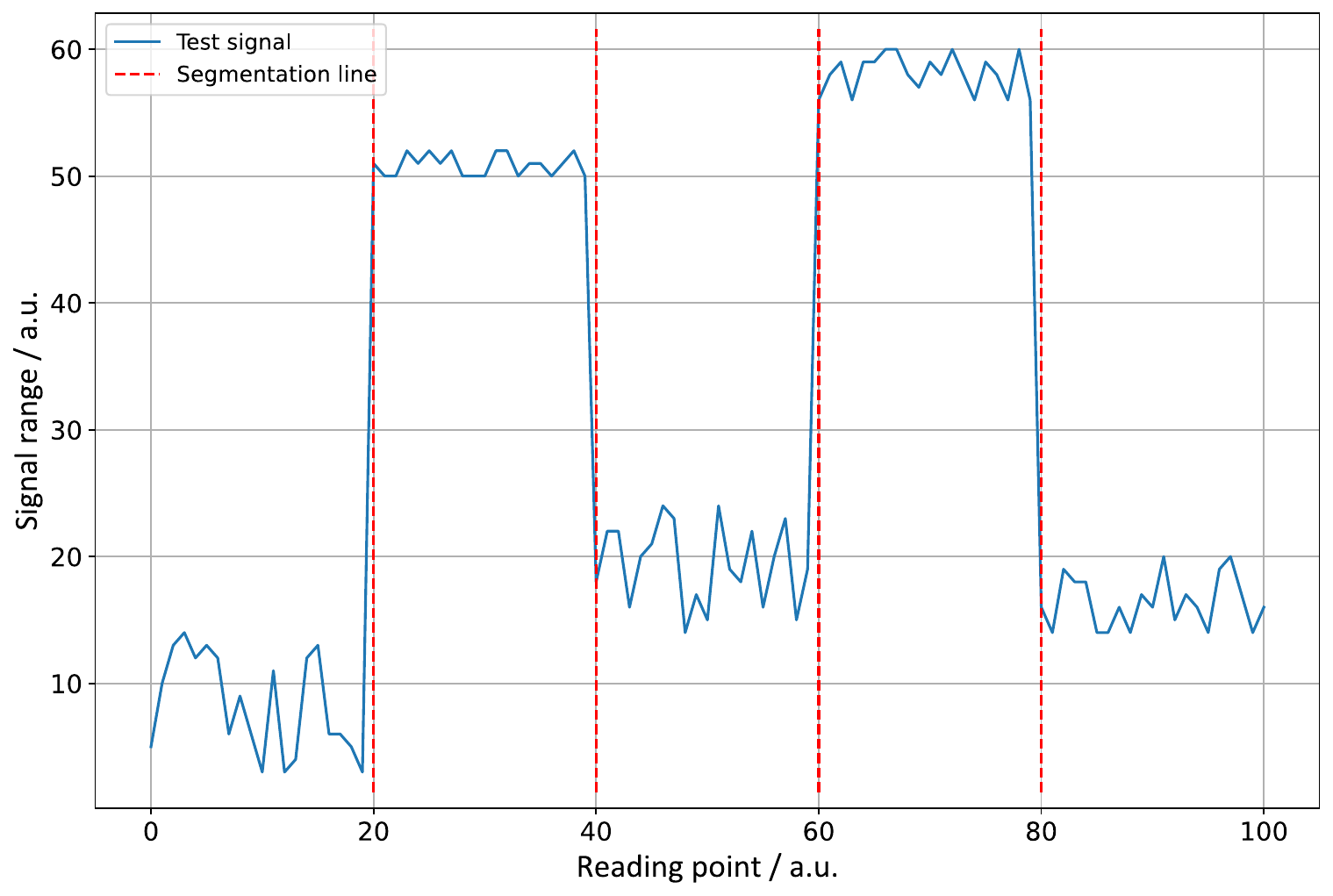}}
			\caption{Test signal with the optimal variance based segmentation.}
			\label{fig:figureVIII}
		\end{figure}

		After calculating the sum of variances for all segments with all possible combinations of segmentations, the optimal solution can be obtained, as seen in figure \ref{fig:figureVIII}.
		
		The time complexity of the brute force program is $\mathcal{O}(n^{m - 1})$, with n being the number of samples in the data set and m being the predefined number of segments. For the test signal roughly 3 hours were needed to calculate the optimal solution. The optimal solution will be used as a reference for the next subsection, which will discuss the use of a custom algorithm for approximating a solution for the segmentation problem.
		\bigskip
		
		\subsection{Custom segmentation algorithm}
		Due to the severe time complexity of calculating the optimal solution using a brute force approach, a custom algorithm was written for approximating a solution for an optimal segmentation of a data set with a fixed number of segments. The algorithm was written in Python and can be denoted with the high-level description shown in the listing below.
		
		\begin{lstlisting}[language=Python, caption=High-level description of the algorithm]
		while new_optimum == True:
			new_optimum = False
			for segment in data_set:
				tempvar = variance(segment) +	|\Suppressnumber|
					variance(segment + 1)	|\Reactivatenumber{5}|
				for data_point in segment:
					tempvar2 = variance(	|\Suppressnumber|
						segment_new_boundary) + 
						variance(segment_new_bondary + 1)	|\Reactivatenumber{7}|
					if tempvar2 < tempvar:
						tempvar = tempvar2
						segment = segment_new_boundary
						new_optimum = True	|\Suppressnumber|
		\end{lstlisting}
		
\begin{figure}[!htbp]
			\centerline{\includegraphics[width=\linewidth]{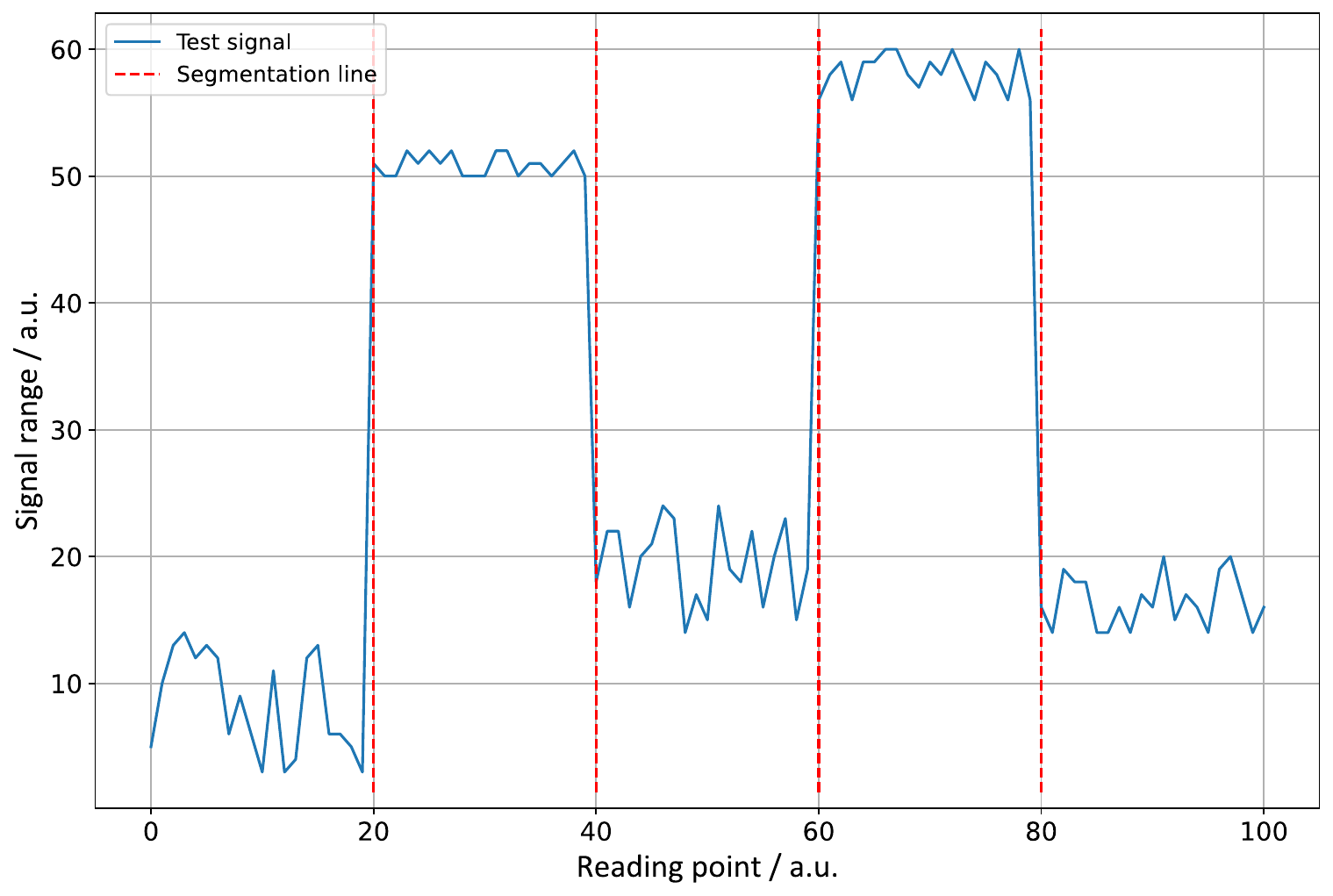}}
			\caption{Test signal with an approximated segmentation from the custom algorithm.}
			\label{fig:figureIX}
		\end{figure}

		After applying the algorithm, the segmentation seen in figure \ref{fig:figureIX} can be obtained.
		
		The algorithm finishes the segmentation of the test signal in under a minute. The time complexity is approximately $\mathcal{O}(n \cdot (m - 1))$. This is a significant improvement compared to the brute force calculation which took 3 hours to generate the solution. Additionally, in the above case, the local optimum calculated with the custom algorithm equals the global optimum. 
		
\begin{figure}[!htbp]
			\centerline{\includegraphics[width=\linewidth]{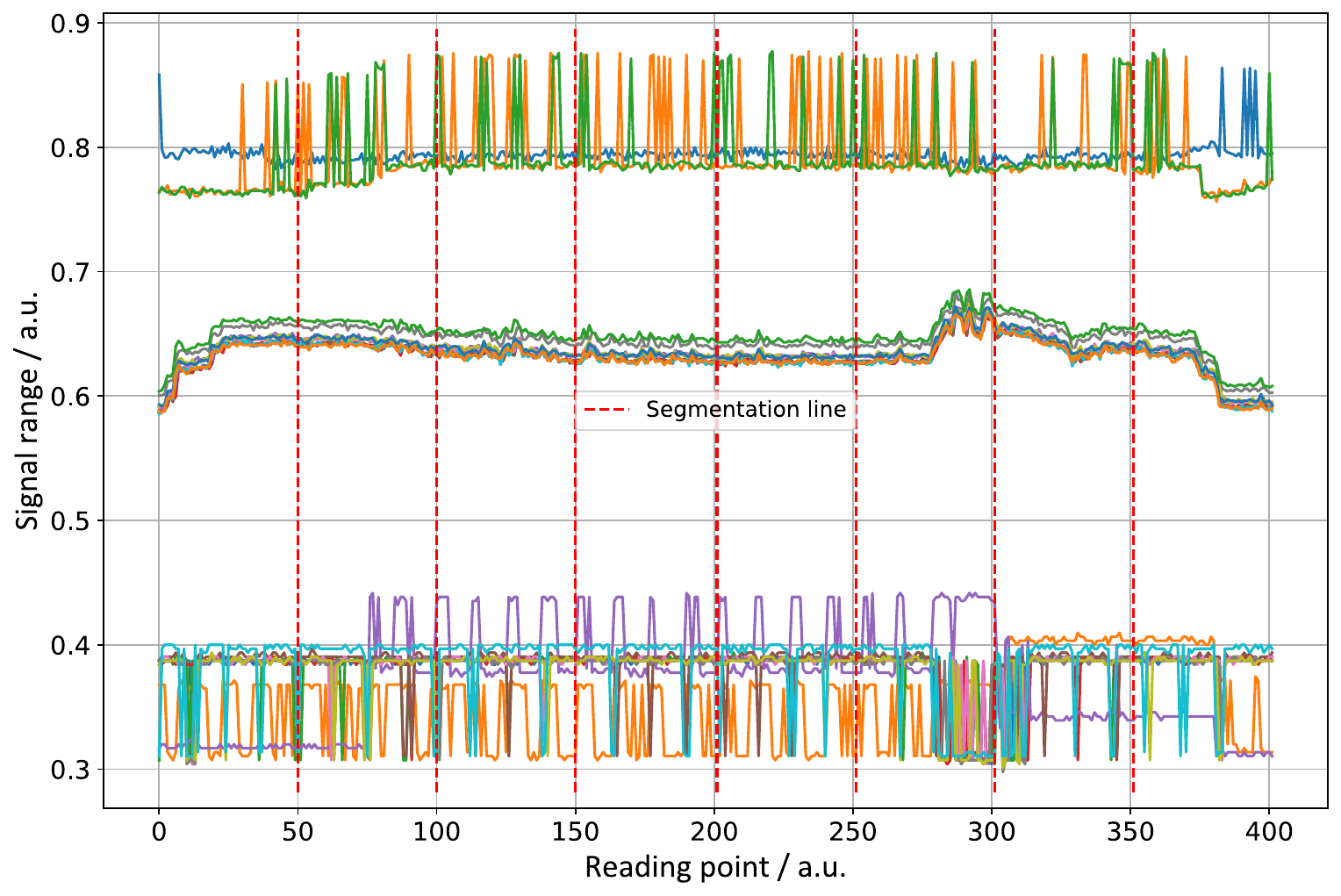}}
			\caption{Filtered signals with an arbitrary presegmentation.}
			\label{fig:figureX}
		\end{figure}

		Following above approach, the custom algorithm is used to calculate an approximate solution for a segmentation of the data set containing the 23 filtered signals. Again, the filtered processes are passed to the algorithm with an arbitrary presegmentation as given in figure \ref{fig:figureX}.
		
\begin{figure}[!htbp]
			\centerline{\includegraphics[width=\linewidth]{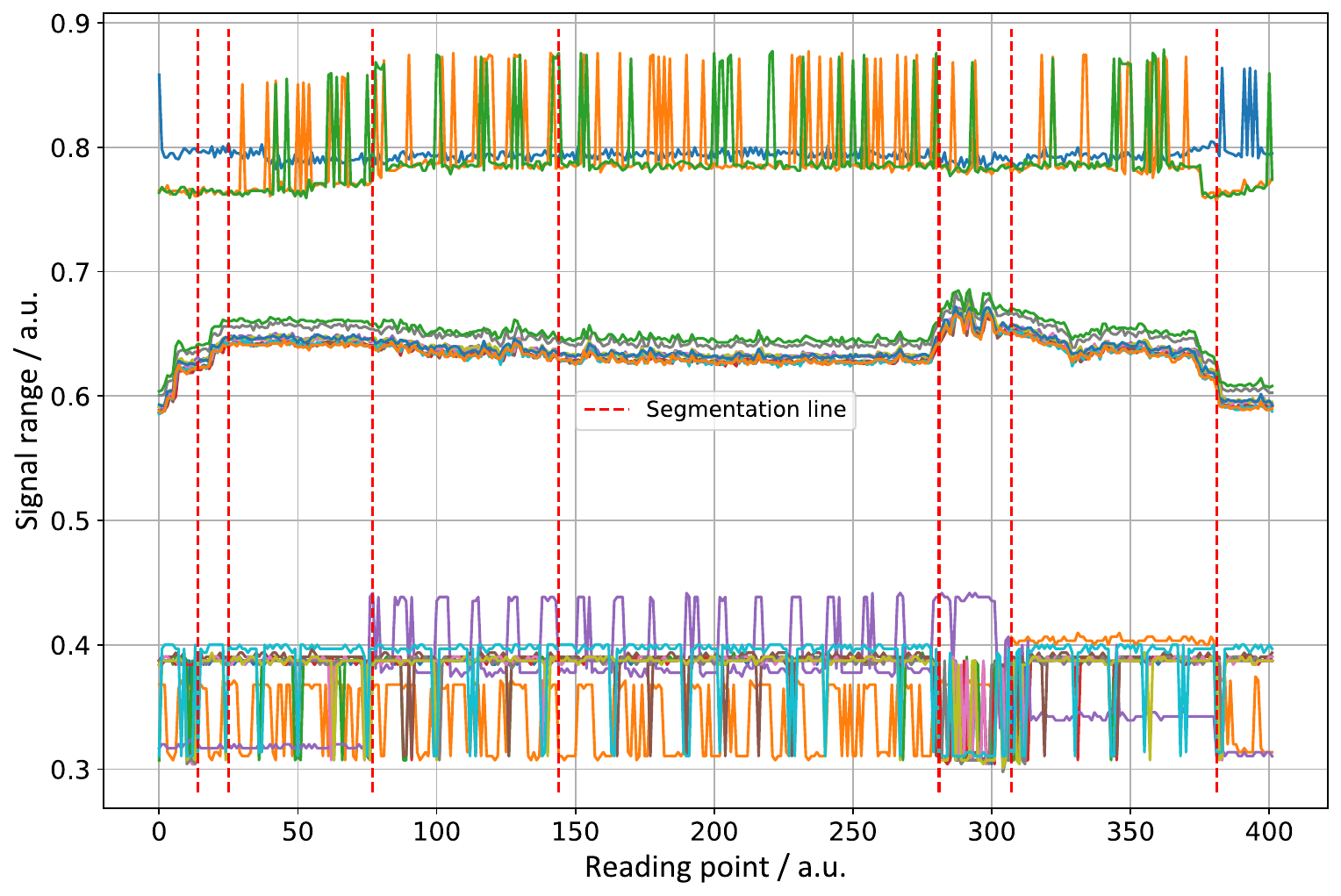}}
			\caption{Filtered signals with an approximated segmentation from the custom algorithm.}
			\label{fig:figureXI}
		\end{figure}

		After applying the algorithm, the solution shown in figure \ref{fig:figureXI} can be obtained.
		
		Especially the sixth and seventh segment in above plot show some interesting details. In the sixth segment, some of the processes in the signal range between approximately 0.3 and 0.4 show significant movement. The teal process drops to 0.3 and remains at that signal range. The other processes show significant fluctuation, different to their movements in the previous segment. The processes in the signal range between 0.6 and 0.7 show a slight rise and fall in the sixth segment not seen in the previous segment. In the signal range between approximately 0.3 and 0.4 and within the seventh segment, the purple process drops to approximately 0.35, the orange process rises to 0.4 and the other processes show similar fluctuations. 	
		\bigskip

	\section{Results}
	In the test signal, a relevant segmentation could be found using the minimization of the variance of each segment of the signal. In our view, this metric is applicable for separating different internal behaviors. This can be seen very clearly in the segmentation of the test signal in figure \ref{fig:figureVIII}, where segments one, three, and five contain data from the data range between 0 and 30 and segments two and four data from the data range between 50 and 60. Sharp changes in a signal sequence, such as the one between the first and second segment in figure \ref{fig:figureVIII}, are captured with the algorithm, as the variance is minimized by including data points from similar data ranges.
	
	The segmentation of the filtered signals shown in figure \ref{fig:figureXI} shows similar characteristics. Each segment contains certain behaviors of the processes not present in the neighboring segments. Sharp changes in the signal sequences, such as between the fifth and sixth segment, are captured by cutting the data with a segmentation line. In terms of variance, the overall variance of the real data set containing the filtered signals was reduced from approximately 0.20 to 0.19. With the test signal, the variance dropped from approximately 1200 to 140 after applying the segmentation algorithm. The small change in variance in the real data set compared to the test signal stems from the significantly higher number of processes in the real data set. The fact is that small changes in one process do not contribute to a significant reduction of variance if there are more processes involved which show no changes in behavior. 
	\bigskip

	\section{Conclusion and future work}
	In this work we motivated the problem of scenario extraction by data segmentation and divided it into the subproblems of finding the number of segments in a data set and the precise segmentation indices.
	We presented an algorithm that gives an approximate solution to the subproblem of finding suitable segmentation indices with a given amount of segments.
	We showed results of the solution by applying it to a test signal and the data set taken out of industry data from our partner. The chosen algorithm is capable of finding good segmentation indices even for multivariate data. The second subproblem of finding a suitable amount of segments is still open and will be investigated in future. Also, additional work like the classification of segments can be seen as a follow-up to this work. 
	\bigskip

	\renewcommand{\bibname}{References}
	\printbibliography
	
\end{document}